\documentclass[a4paper,twocolumn,11pt,unpublished,allowtoday]{quantumarticle}
\pdfoutput=1

\usepackage{amsmath,amssymb,amsfonts}
\usepackage{bm}
\usepackage{graphicx}
\usepackage{xcolor}
\usepackage{array}
\usepackage{physics}
\usepackage{algorithm}
\usepackage{algpseudocode}
\usepackage{placeins}
\usepackage[numbers,sort&compress]{natbib}
\usepackage{hyperref}

\begin{document}

\title{Finite-temperature quantum Krylov method from real-time overlaps}

\author{Hiroto Yamamoto}
\email{yamamoto.hiroto.r8@dc.tohoku.ac.jp}
\affiliation{Department of Physics, Tohoku University, Sendai, Miyagi 980-8578, Japan}

\author{Katsuhiro Morita}
\email{katsuhiro.morita.c7@mso.tohoku.ac.jp}
\affiliation{Department of Physics, Tohoku University, Sendai, Miyagi 980-8578, Japan}

\date{\today}

\begin{abstract}
Accurately evaluating finite-temperature properties of quantum many-body systems remains a central challenge. Many existing quantum approaches require thermal-state preparation at each target temperature, making low-temperature calculations especially demanding in terms of circuit depth and accuracy.

In this work, we present the finite-temperature quantum Krylov (FTQK) method, which combines stochastic trace estimation with a Hermitian generalized eigenvalue problem based on $\cos(\tilde H)$. The only quantum-side input required is the temperature-independent real-time overlap sequence $g_n=\langle\phi|e^{-in\tau H}|\phi\rangle$. All benchmark results presented in this work were obtained from classical numerical simulations of the quantum-circuit protocol. For periodic spin-$\frac{1}{2}$ Heisenberg chains with $N=14$ and $18$, FTQK accurately reproduces the specific heat, magnetic susceptibility, and entropy in the noiseless case. It also accurately reproduces the specific heat of the $N=14$ honeycomb Kitaev model, demonstrating that FTQK can be applied beyond simple one-dimensional chains to a model on a two-dimensional lattice with bond-dependent interactions. Furthermore, in a proof-of-principle finite-shot study for the $N=14$ Heisenberg chain, the main thermodynamic features remain well preserved even at $\sigma=10^{-3}$. Here, 
$\sigma=1/\sqrt{N_{\mathrm{shot}}}$, where $N_{\mathrm{shot}}$ is the number of shots used to estimate  an overlap. These results demonstrate the potential of FTQK for finite-temperature calculations using reusable, temperature-independent real-time overlap data.

\end{abstract}

\maketitle

\section{Introduction}
\label{sec:introduction}

Elucidating finite-temperature properties of quantum many-body systems is important for understanding quantum phenomena realized in materials. In particular, finite-temperature behavior is indispensable for understanding collective quantum phenomena such as superconducting transitions and quantum spin liquids. Thermodynamic quantities such as the specific heat and magnetic susceptibility directly reflect these phenomena and thus serve as key observables linking theory and experiment.

However, finite-temperature quantum many-body problems remain notoriously challenging in general. Although a variety of classical approaches have been developed~\cite{Schollwock2011DMRG}, quantitatively evaluating thermodynamic quantities down to low temperatures remains challenging in the presence of frustration, high dimensionality, or the sign problem~\cite{Troyer2005SignProblem}. The finite-temperature Lanczos method (FTLM) provides a framework for accurately evaluating finite-temperature thermodynamic quantities within a Krylov subspace~\cite{Jaklic1994FTLM,Jaklic2000FTLMReview,Schnack2018Kagome42,Schnack2020FTLMAccuracy,Morita2020FTLM,Morita2022OFTLM,Morita2023KagomeOFTLM}. However, the accessible system sizes remain strongly limited by the exponential growth of the Hilbert-space dimension. Establishing computational principles for evaluating thermodynamic quantities therefore remains an important challenge for both classical and quantum computation~\cite{Chen2025EfficientThermal}.

In recent years, various approaches have been proposed for finite-temperature calculations on quantum computers, including quantum imaginary-time evolution~\cite{Motta2020QITE,Gomes2021AVQITE,Nishi2021QITENISQ,YeterAydeniz2022QITEQFT,Kamakari2022OpenQITE,Tsuchimochi2023ImprovedQITE,Nishi2023OptimalPITE,Hejazi2024AdiabaticQITE,Cianci2024SSQITE,Nishi2025EncodedPITE}, methods based on thermal pure quantum states~\cite{PhysRevA.108.012404,PhysRevLett.131.081901}, and approaches based on quantum Gibbs sampling~\cite{Chen2025EfficientThermal}. While these approaches mark important progress toward finite-temperature quantum simulation, many of them take the target inverse temperature $\beta$ as input and aim to construct on a quantum device a thermal state, or an approximate one, at that temperature. As a result, quantum computations generally depend explicitly on the target temperature, making thermal-state preparation or temperature-dependent state updates a practical bottleneck, particularly in the low-temperature regime where circuit depths can become large and the accuracy can deteriorate~\cite{PhysRevB.108.085102,Ye2025QKFE}.

Quantum Krylov-subspace methods provide a complementary route in which spectral information is reconstructed from a low-dimensional subspace generated by quantum dynamics. A range of approaches based on Hamiltonian, unitary, or real-time Krylov constructions have been developed for ground- and excited-state estimation~\cite{Stair2020MRSQK,Seki2021QuantumPowerMethod,Baker2021LanczosRecursion,Cohn2021QFD,Cortes2022QKSD,Kirby2023ExactLanczos,Shen2023RealTimeKrylov}. Related developments include stochastic, fast-forwarding, variational, Davidson-type, block-Lanczos, measurement-efficient, and fragment-based extensions~\cite{Cortes2022FastForwarding,Stair2023StochasticQK,Filip2024VPEVFF,Tkachenko2024QuantumDavidson,Baker2024BlockLanczos,Zhang2024MeasurementEfficientQKSD,DCunha2024FragmentInit}. Recent applications and variants have further extended quantum Krylov ideas to lattice-gauge, molecular, and many-body problems~\cite{Anderson2025LGTQK,Oumarou2025MolecularPropertiesQKSD,Yoshioka2025KrylovProcessor,Shen2025EstimatingEigenenergies,Misciasci2026SKQDHeisenberg}. The conditioning of the generalized eigenvalue problem and its sensitivity to sampling errors have also been analyzed in detail~\cite{Epperly2022TheoryQSD,Kirby2024QKErrors,Lee2024SamplingErrorQKSD,Lee2025ReducingSamplingError}. More recently, the finite-temperature quantum Krylov method (FTQK) studied here was independently and contemporaneously introduced under the name Quantum Finite Temperature Lanczos Method (QFTLM)~\cite{Gentinetta2026QFTLM}. The terms QFTLM and FTQK therefore refer to the same method; the distinction is purely terminological. The two studies nevertheless differ in the projected operator used in the generalized eigenvalue problem, the benchmark models considered, and the treatment of finite-shot noise, as discussed in Sec.~\ref{subsec:related_methods}.

Here we present the formulation of FTQK used in this work, in which finite-temperature thermodynamic quantities are reconstructed from unitary real-time evolution that can be naturally implemented on quantum circuits. The only quantum-side input required by the protocol is the real-time overlap sequence

\begin{equation}
g_n = \langle \phi| e^{-in\tau H} |\phi \rangle
\qquad (n\in\mathbb{Z}),
\label{eq:intro_gn}
\end{equation}

associated with the unitary real-time evolution under the Hamiltonian $H$. Because these overlaps are obtained without specifying a target temperature, the same data can be used in classical post-processing to reconstruct thermodynamic quantities over a broad temperature range. Our method shares with FTLM the basic idea of extracting finite-temperature quantities from a Krylov subspace, but differs in that its quantum input is described by overlaps generated by unitary real-time evolution rather than by repeated classical applications of $H$.

In this work, we benchmark FTQK for the one-dimensional spin-$\frac{1}{2}$ Heisenberg model with periodic boundary conditions (PBC). All benchmark results presented in this work are obtained from classical numerical simulations of the quantum-circuit protocol. For $N=14$ and $N=18$, the noiseless FTQK results for the specific heat, magnetic susceptibility, and entropy agree closely with those obtained by full exact diagonalization. For $N=18$, we further examine the dependence on the Krylov dimension $D$ and observe systematic convergence toward the exact result as $D$ increases. For $N=14$, we perform a proof-of-principle finite-shot study by introducing statistical fluctuations into the real-time overlaps $g_n$; with the stabilization procedures employed here, the main thermodynamic features remain well preserved up to $\sigma=10^{-3}$. We also present a noiseless benchmark of the specific heat for the $N=14$ honeycomb Kitaev model, demonstrating that FTQK is applicable beyond simple one-dimensional chains to a model on a two-dimensional lattice with bond-dependent interactions.

The spectral reconstruction is performed through a Hermitian generalized eigenvalue problem based on $\cos(\tilde H)$, together with an affine spectral mapping that allows approximate energies to be recovered through the inverse cosine. In the Heisenberg-chain benchmark, symmetry-preserving pseudorandom states prepared in fixed $S_{\mathrm{tot}}^z$ sectors enable sector-resolved trace estimation and the evaluation of the magnetic susceptibility, while the real-time evolution itself is performed in the full Hilbert space.

The finite-shot calculations considered here model statistical fluctuations through a Gaussian sampling model and do not include device-specific effects such as gate errors, readout errors, or decoherence. Direct validation on quantum hardware is left for future work.

The remainder of this article is organized as follows. Section~\ref{sec:ftqk} introduces the finite-temperature trace-estimation framework, real-time Krylov formulation, and thermodynamic estimators. Section~\ref{sec:stabilized} describes the affine mapping, the Hermitian generalized eigenvalue problem based on $\cos(\tilde H)$ and its regularization. Section~\ref{sec:implementation} describes the benchmark setup for the Heisenberg chain, including the model, Trotterized real-time evolution, pseudorandom-state preparation, and finite-shot model. Section~\ref{sec:benchmarks} presents the Heisenberg-chain benchmarks, including the Krylov-dimension dependence and finite-shot effects, together with a noiseless benchmark for the honeycomb Kitaev model. Section~\ref{sec:discussion} discusses the relation to existing methods, limitations, and outlook. Section~\ref{sec:conclusion} summarizes the main conclusions. Technical details and supplementary checks are collected in Appendices~\ref{app:circuit}--\ref{app:additional_model}.

\section{Finite-temperature quantum Krylov method}
\label{sec:ftqk}

The FTQK method is based on stochastic trace estimation combined with a Krylov subspace generated by unitary real-time evolution. When the Hilbert space is decomposed into symmetry sectors labeled by a conserved quantum number $q$, the trace of an operator $A$ restricted to sector $q$ can be approximated as
\begin{equation}
\mathrm{Tr}_q A
\approx
\frac{N_{\mathrm{st}}^{(q)}}{R}
\sum_{r=1}^{R}
\langle \phi_0^{(r,q)}|A|\phi_0^{(r,q)}\rangle ,
\label{eq:trace_estimator}
\end{equation}
where $N_{\mathrm{st}}^{(q)}$ is the Hilbert-space dimension of sector $q$, $R$ is the number of sampled initial states, and $|\phi_0^{(r,q)}\rangle$ is the $r$th normalized random state in that sector. In the Heisenberg-chain benchmark introduced in Sec.~\ref{sec:implementation}, $q$ is identified with the conserved total magnetization $S^z_{\mathrm{tot}}$. The finite-depth circuit states used in the numerical implementation are pseudorandom rather than exact Haar-random states; their construction is described in Sec.~\ref{subsec:pseudorandom} and Appendix~\ref{app:circuit}.

\subsection{Real-time Krylov formulation}
\label{subsec:realtime_krylov}

For each $|\phi_0^{(r,q)}\rangle$, we define the real-time overlap sequence
\begin{equation}
g_n^{(r,q)}
=
\langle \phi_0^{(r,q)}| e^{-in\tilde{H}} |\phi_0^{(r,q)}\rangle,
\label{eq:gn}
\end{equation}
which satisfies $g_{-n}^{(r,q)}=\left(g_n^{(r,q)}\right)^*$ because $\tilde{H}$ is Hermitian. Here $\tilde H$ denotes the affine-transformed Hamiltonian used in the practical implementation; its relation to $H$ is given in Sec.~\ref{subsec:affine}. Such overlaps can be estimated on quantum hardware by standard interferometric methods such as the Hadamard test.

Writing $U=e^{-i\tilde{H}}$, we introduce the real-time Krylov subspace
\begin{equation}
\begin{aligned}
\mathcal{K}_D(U,|\phi_0^{(r,q)}\rangle)
={}&\mathrm{span}\{ |\phi_0^{(r,q)}\rangle,
U|\phi_0^{(r,q)}\rangle,\dots,\\
&\hspace{4.0em}U^{D-1}|\phi_0^{(r,q)}\rangle \}, 
\end{aligned}
\label{eq:krylov_subspace}
\end{equation}
and the basis states $|\phi_n^{(r,q)}\rangle=U^n|\phi_0^{(r,q)}\rangle$ $(n=0,1,\dots,D-1)$. The overlap matrix then takes the Toeplitz form
\begin{equation}
S_{nn'}^{(r,q)}
=
\langle \phi_n^{(r,q)}|\phi_{n'}^{(r,q)}\rangle
=
g_{n'-n}^{(r,q)}.
\label{eq:S}
\end{equation}

In the practical Hermitian formulation adopted here, we project
\begin{equation}
F=\cos(\tilde H)=\frac{U+U^{-1}}{2}.
\label{eq:F_operator}
\end{equation}
Its matrix elements are written entirely in terms of the same real-time overlaps,
\begin{equation}
\begin{aligned}
F_{nn'}^{(r,q)}
={}&\langle \phi_n^{(r,q)}| \cos(\tilde{H}) |\phi_{n'}^{(r,q)}\rangle\\
={}&\frac{g_{n'-n+1}^{(r,q)}+g_{n'-n-1}^{(r,q)}}{2}.
\end{aligned}
\label{eq:F}
\end{equation}
Thus, no separate measurement of Hamiltonian matrix elements is required to construct this generalized eigenvalue problem once the overlap sequence has been obtained.

We then solve the generalized eigenvalue problem
\begin{equation}
F^{(r,q)}\boldsymbol{u}_{j}^{(r,q)}
=
\lambda_{j}^{(r,q)}S^{(r,q)}\boldsymbol{u}_{j}^{(r,q)}.
\end{equation}
Here, \(j\) labels the generalized eigenpairs, \(\lambda_{j}^{(r,q)}\) are approximate eigenvalues of
\(\cos(\widetilde{H})\), and \(\boldsymbol{u}_{j}^{(r,q)}\)
are the corresponding generalized eigenvectors.
 From these, we reconstruct approximate energies as described in Sec.~\ref{subsec:affine}, and define the corresponding approximate eigenstates as
\begin{equation}
|\psi_j^{(r,q)}\rangle
=
\sum_{n=0}^{D-1}u_{nj}^{(r,q)}|\phi_n^{(r,q)}\rangle.
\label{eq:ritz_state}
\end{equation}
The corresponding weight is
\begin{equation}
w_j^{(r,q)}
=
\left|
\langle \phi_0^{(r,q)}|\psi_j^{(r,q)}\rangle
\right|^2,
\label{eq:weight}
\end{equation}
with each mode normalized such that
$\langle\psi_j^{(r,q)}|\psi_j^{(r,q)}\rangle=1$.

The quantum input in Eqs.~\eqref{eq:gn}--\eqref{eq:F} contains no target temperature. Once the overlap data have been collected and the Ritz information has been reconstructed, the same data can be reused in the classical post-processing for any desired $\beta=1/T$.

\subsection{Thermodynamic estimators}
\label{subsec:thermo_estimators}

When evaluating finite-temperature thermodynamic quantities using FTQK, the partition function is given by the following expression, assuming zero magnetic field and $k_B=1$:
\begin{equation}
Z(T)
\approx
\sum_{q=-M_{\mathrm{sat}}}^{M_{\mathrm{sat}}}
\frac{N_{\mathrm{st}}^{(q)}}{R}
\sum_{r=1}^{R}
\sum_{j=0}^{D_{\mathrm{eff}}^{(r,q)}-1}
e^{-\beta E_{j}^{(r,q)}}
\, w_{j}^{(r,q)} .
\label{eq:Z}
\end{equation}
The corresponding approximate energies \(E_j^{(r,q)}\) are reconstructed as described in Sec.~\ref{subsec:affine}.
For the magnetic-sector decomposition used in the Heisenberg-chain benchmark, $q$ denotes the total-magnetization quantum number and $M_{\mathrm{sat}}$ its maximum absolute value. $N_{\mathrm{st}}^{(q)}$ is the dimension of sector $q$, $D_{\mathrm{eff}}^{(r,q)}$ is the effective Krylov dimension after the thresholding procedure described in Sec.~\ref{subsec:regularization}, and $R$ is the number of samples for the initial state $|\phi_0^{(r,q)}\rangle$.

The internal energy is estimated as
\begin{equation}
\begin{aligned}
E(T)
\approx{}&\frac{1}{Z(T)}
\sum_{q=-M_{\mathrm{sat}}}^{M_{\mathrm{sat}}}
\frac{N_{\mathrm{st}}^{(q)}}{R}
\sum_{r=1}^{R}\\
&\times\sum_{j=0}^{D_{\mathrm{eff}}^{(r,q)}-1}
E_{j}^{(r,q)} e^{-\beta E_{j}^{(r,q)}} w_{j}^{(r,q)} .
\end{aligned}
\label{eq:E_thermo}
\end{equation}
The magnetization is
\begin{equation}
\begin{aligned}
M(T)
\approx{}&\frac{1}{Z(T)}
\sum_{q=-M_{\mathrm{sat}}}^{M_{\mathrm{sat}}}
\frac{N_{\mathrm{st}}^{(q)}}{R}
\sum_{r=1}^{R}\\
&\times\sum_{j=0}^{D_{\mathrm{eff}}^{(r,q)}-1}
q\, e^{-\beta E_{j}^{(r,q)}} w_{j}^{(r,q)} .
\end{aligned}
\label{eq:M_thermo}
\end{equation}
Accordingly, the specific heat is
\begin{equation}
\begin{aligned}
C(T)
\approx{}&\frac{1}{T^2}\Biggl[
\frac{1}{Z(T)}
\sum_{q=-M_{\mathrm{sat}}}^{M_{\mathrm{sat}}}
\frac{N_{\mathrm{st}}^{(q)}}{R}
\sum_{r=1}^{R}\\
&\times\sum_{j=0}^{D_{\mathrm{eff}}^{(r,q)}-1}
\left(E_{j}^{(r,q)}\right)^2
e^{-\beta E_{j}^{(r,q)}} w_{j}^{(r,q)}\\
&-\left(E(T)\right)^2\Biggr] .
\end{aligned}
\label{eq:C_thermo}
\end{equation}
The entropy is
\begin{equation}
S_\mathrm{m}(T)
\approx
\frac{E(T)}{T}
+
\ln Z(T),
\label{eq:S_thermo}
\end{equation}
and the magnetic susceptibility is
\begin{equation}
\begin{aligned}
\chi(T)
\approx{}&\frac{1}{T}\Biggl[
\frac{1}{Z(T)}
\sum_{q=-M_{\mathrm{sat}}}^{M_{\mathrm{sat}}}
\frac{N_{\mathrm{st}}^{(q)}}{R}
\sum_{r=1}^{R}\\
&\times\sum_{j=0}^{D_{\mathrm{eff}}^{(r,q)}-1}
q^2 e^{-\beta E_{j}^{(r,q)}} w_{j}^{(r,q)}
-\left(M(T)\right)^2\Biggr] .
\end{aligned}
\label{eq:chi_thermo}
\end{equation}
These expressions preserve the sector-dimension factor $N_{\mathrm{st}}^{(q)}$, the average over $R$ samples, and the reconstructed spectral weights.

\section{Stabilized \texorpdfstring{$\cos(\tilde H)$}{cos(H)}-based formulation}
\label{sec:stabilized}

FTQK denotes the overall finite-temperature framework described above. The $\cos(\tilde H)$-based generalized eigenvalue problem is the practical Hermitian formulation adopted in the present numerical study. This section describes the affine spectral mapping used to make the inverse cosine single valued and the linear-algebra regularization used when the overlap matrix becomes ill-conditioned.
Throughout this section, unless otherwise stated, we consider a fixed sample $r$ and symmetry sector $q$ and suppress the labels $(r,q)$ on the Krylov-space quantities for notational simplicity.

\subsection{Affine spectral mapping and energy reconstruction}
\label{subsec:affine}

We first apply an affine transformation to the original Hamiltonian,
\begin{equation}
\tilde{H} = \tau^{(q)} H + \theta^{(q)} I,
\label{eq:Htilde}
\end{equation}
where the constants $\tau^{(q)}$ and $\theta^{(q)}$ are pre-computed classically for each symmetry sector $q$. The purpose of this transformation is to confine a reference spectral interval to $[0,\pi]$, so that the approximate energy eigenvalues can be recovered using the inverse cosine function.

As the transformed energy eigenvalues approach $0$ or $\pi$, the derivative of the $\arccos$ function diverges, leading to a degradation in resolution. Conversely, if we set an excessively large margin to avoid this divergence---thereby overly narrowing the transformed spectral width---the Krylov basis becomes highly susceptible to linear dependence.

We first pre-evaluate the minimum and maximum energies of the Hamiltonian $H$ in sector $q$ to define the reference spectral interval
$[E_{\min,\mathrm{ref}}^{(q)},E_{\max,\mathrm{ref}}^{(q)}]$.
We then map this interval into
$[\Delta_{\min},\Delta_{\max}]$ with
\begin{equation}
\tau^{(q)}
=
\frac{\Delta_{\max}-\Delta_{\min}}
{E^{(q)}_{\max,\mathrm{ref}}-E^{(q)}_{\min,\mathrm{ref}}},
\label{eq:tau_affine}
\end{equation}
and
\begin{equation}
\theta^{(q)}
=
-\tau^{(q)} E^{(q)}_{\min,\mathrm{ref}}
+
\Delta_{\min}.
\label{eq:theta_affine}
\end{equation}

In the present benchmarks, we set \(\Delta_{\min}=0\) and \(\Delta_{\max}=\pi\). For the Heisenberg-chain benchmarks, the spectral upper bound can be evaluated exactly, and we therefore set $E_{\max,\mathrm{ref}}^{(q)}=E_{\max,\mathrm{exact}}^{(q)}$. For the lower bound, using the exact minimum energy directly as the reference value would map the lowest-energy state to $\Delta_{\min}=0$, where the inverse-cosine resolution deteriorates. Therefore, we introduce a relative margin $r_{\mathrm{low}}$ and define
\begin{equation}
E_{\min,\mathrm{ref}}^{(q)}
=
E_{\min,\mathrm{exact}}^{(q)}
-
\left|E_{\min,\mathrm{exact}}^{(q)}\right|\, r_{\mathrm{low}} .
\label{eq:Eref}
\end{equation}
This intentionally sets the reference lower bound below the exact lower bound, mapping the actual spectral lower bound to a value strictly greater than $0$ in the transformed interval.

For both $N=14$ and $N=18$, we set $r_{\mathrm{low}}=0.05$.
Even when full exact diagonalization is intractable on classical computers, the present method requires only conservative sector-resolved spectral bounds for the affine mapping. 
For antiferromagnetic Heisenberg models, the upper edge of the spectrum can be determined analytically and exactly for arbitrary system size.
A conservative lower bound can likewise be obtained using analytical lower-bound methods for antiferromagnetic spin systems~\cite{Anderson1951,Tarrach1990}.
Based on this perspective, we adopted the above choice of $r_{\rm low}$.

The overlap sequence for $\tilde H$ is related to evolution under the original Hamiltonian by
\begin{equation}
\langle \phi_0 | e^{-in \tilde{H}} | \phi_0 \rangle
=
e^{-in\theta^{(q)}}
\langle \phi_0 | e^{-in\tau^{(q)} H} | \phi_0 \rangle .
\label{eq:affine_overlap}
\end{equation}
Therefore, we only need to perform the overlap measurement for the time-evolution operator $e^{-in\tau^{(q)}H}$ and multiply the obtained result by the known phase factor $e^{-in\theta^{(q)}}$ on a classical computer.

If $\tilde{H}|\Psi_k\rangle=\tilde{E}_k|\Psi_k\rangle$, then
$\cos(\tilde{H})|\Psi_k\rangle=\cos(\tilde{E}_k)|\Psi_k\rangle$.
Because $\arccos(x)$ is single valued on the principal branch
$\tilde E_k\in[0,\pi]$, the eigenvalues
$\lambda_k=\cos(\tilde E_k)$ determine
$\tilde E_k=\arccos(\lambda_k)$.
Numerical errors or finite-shot noise may nevertheless yield
$\lambda>1$ or $\lambda<-1$. We therefore clamp
\begin{equation}
\lambda_{\mathrm{clamp}}
=
\min\!\bigl(1,\max(-1,\lambda)\bigr)
\label{eq:lambda_clamp}
\end{equation}
and recover the approximate energy of the original Hamiltonian as
\begin{equation}
E^{(q)}
=
\frac{\arccos(\lambda_{\mathrm{clamp}})-\theta^{(q)}}{\tau^{(q)}}.
\label{eq:E_recover_final}
\end{equation}

\subsection{Regularized generalized eigenvalue problem}
\label{subsec:regularization}

Due to the linear dependence of the Krylov basis or numerical errors, the overlap matrix $S$ may have eigenvalues close to zero, leading to ill-conditioning. To avoid this, we perform regularization by introducing a threshold $\varepsilon$ and discarding the subspace corresponding to these small eigenvalues. Specifically, we first perform an eigenvalue decomposition of the Hermitian matrix $S$ with Krylov dimension $D$ and sort the eigenvalues in descending order $(s_1\ge s_2\ge\dots\ge s_D)$:
\begin{equation}
S = W \Sigma W^\dagger,
\qquad
\Sigma=\mathrm{diag}(s_1,s_2,\dots,s_D).
\label{eq:S_evd}
\end{equation}
Here, $W=(\mathbf{w}_1,\dots,\mathbf{w}_D)$ is a $D\times D$ unitary matrix whose columns are the eigenvectors. Assuming there are $D_{\mathrm{eff}}$ effective eigenvalues not smaller than the threshold $\varepsilon$, i.e.,
$s_{D_{\mathrm{eff}}}\ge\varepsilon> s_{D_{\mathrm{eff}}+1}$,
we define
\begin{equation}
\Sigma_\varepsilon
=
\mathrm{diag}(s_1,\dots,s_{D_{\mathrm{eff}}})
\label{eq:Sigma_eps}
\end{equation}
and
\begin{equation}
X_\varepsilon
=
W_\varepsilon\Sigma_\varepsilon^{-1/2}.
\label{eq:X_eps}
\end{equation}
This satisfies
$X_\varepsilon^\dagger S X_\varepsilon=I_{D_{\mathrm{eff}}}$.
Therefore, by applying the variable transformation
$\mathbf v=X_\varepsilon\mathbf y$
to the generalized eigenvalue problem
$F\mathbf v=\lambda S\mathbf v$,
we obtain the $D_{\mathrm{eff}}$-dimensional standard eigenvalue problem
\begin{equation}
F_\varepsilon\mathbf y
=
\lambda\mathbf y,
\qquad
F_\varepsilon
\equiv
X_\varepsilon^\dagger F X_\varepsilon.
\label{eq:whitened_evp}
\end{equation}
This thresholding is a linear-algebra regularization of the generalized eigenvalue problem.
It is conceptually distinct from the trimmed-mean stabilization applied to finite-shot data, which is discussed in
Sec.~\ref{subsec:shot_stabilization} and detailed in Appendix~B.2.

\subsection{Stabilization under finite-shot noise}
\label{subsec:shot_stabilization}

Under the finite-shot noise model described in Sec.~\ref{subsec:noise_model}, the smallest eigenvalues of the overlap matrix can amplify statistical fluctuations and produce unphysical Ritz values. We therefore solve the generalized eigenvalue problem in a regularized subspace obtained by discarding small-eigenvalue modes of the overlap matrix. 
 The eigenvalue threshold used to discard these modes is selected automatically for each sample.
In addition, for the noisy data only, we apply the trimmed-mean stabilization described in Appendix~\ref{appsub:trimmed_mean}: the reconstructed ground- and first-excited-state energies and their weights are replaced by representative values obtained from the central $50\%$ of the sample distribution.
 The automatic threshold selection and trimmed-mean stabilization are not used in noiseless FTQK calculations but are introduced to stabilize calculations with finite-shot data.
Their detailed implementation is given in Appendix~\ref{app:noise_stability}.

\section{Benchmark setup}
\label{sec:implementation}

\subsection{Heisenberg model}
\label{subsec:heisenberg_model}

For the numerical validation, we consider the one-dimensional spin-$\frac{1}{2}$ Heisenberg model with PBC. The Hamiltonian is given by
\begin{equation}
H
=
J \sum_{i=1}^{N}
\bm{S}_i \cdot \bm{S}_{i+1},
\qquad
\bm{S}_{N+1}\equiv \bm{S}_1 ,
\label{eq:Heisenberg_H}
\end{equation}
where $\bm{S}_i=(S_i^x,S_i^y,S_i^z)$ is the spin-$\frac{1}{2}$ operator at site $i$. Throughout this work, we set $J=1$ and assume an even number of sites $N$.

The Hamiltonian in Eq.~\eqref{eq:Heisenberg_H} has a U(1) symmetry, and the total magnetization
\begin{equation}
S^z_{\mathrm{tot}}=\sum_{i=1}^N S_i^z
\label{eq:Sztot}
\end{equation}
is conserved. We therefore label the magnetization sectors by $S^z_{\mathrm{tot}}=q$. In addition, due to the spin-flip symmetry of Eq.~\eqref{eq:Heisenberg_H}, the eigenstates in the $+q$ and $-q$ sectors are degenerate.
We therefore perform calculations only for $q\geq 0$.
 For compactness, however, the thermodynamic formulas in Sec.~\ref{subsec:thermo_estimators} retain sums over both signs of $q$.

In the present implementation, the real-time evolution is performed in the full $2^N$-dimensional Hilbert space rather than in an explicitly symmetry-reduced basis. The sector restriction is imposed on the initial pseudorandom states. Since $[H,S^z_{\mathrm{tot}}]=0$, a state initialized in sector $q$ remains in the same sector under time evolution. This allows the sector-resolved trace estimation of Sec.~\ref{sec:ftqk} to be combined with a full-Hilbert-space quantum-circuit implementation.

\subsection{Trotterized real-time evolution}
\label{subsec:trotter_evolution}

To implement the time-evolution operator $U(\tau)=e^{-iH\tau}$, we divide the Hamiltonian into even- and odd-bond terms,
\begin{equation}
\begin{aligned}
H_{\mathrm{even}}
&=J\sum_{i\ {\rm even}}\bm S_i\cdot\bm S_{i+1},\\
H_{\mathrm{odd}}
&=J\sum_{i\ {\rm odd}}\bm S_i\cdot\bm S_{i+1}.
\end{aligned}
\label{eq:even_odd}
\end{equation}
Because bond terms with the same parity commute, we use the second-order Suzuki--Trotter decomposition~\cite{Trotter1959,Suzuki1976,Lloyd1996}
\begin{equation}
\begin{aligned}
U(\tau)
={}&\prod_{k=1}^{N_{\mathrm{step}}}
\left(
e^{-i\frac{dt}{2}H_{\mathrm{even}}}
e^{-idtH_{\mathrm{odd}}}
e^{-i\frac{dt}{2}H_{\mathrm{even}}}
\right)\\
&+\mathcal{O}(\tau dt^2).
\end{aligned}
\label{eq:Trotter}
\end{equation}
Here, $\tau$ is divided into $N_{\mathrm{step}}$ Trotter steps, with
$dt=\tau/N_{\mathrm{step}}$.

All finite-temperature results presented in this study are based on simulations with $N_{\mathrm{step}}=1$. 

\subsection{Symmetry-preserving pseudorandom initial states}
\label{subsec:pseudorandom}

The initial state $|\phi_0^{(r,q)}\rangle$ used in the
Heisenberg-chain benchmark is not a Haar-random state belonging to
$S^z_{\rm tot}=q$, but rather a pseudorandom state generated by a
finite-depth random circuit that conserves the total magnetization. In the following, we assume $q\ge0$. First, setting 
$n_\downarrow=N/2-q$ and $n_\uparrow=N/2+q$, we randomly select a computational basis state $|c^{(r,q)}\rangle$ containing $n_\downarrow$ spins in the $\ket{1}$ state and $n_\uparrow$ spins in the $\ket{0}$ state. We then apply $L_{\rm rand}$ layers of a two-qubit $S^z_{\rm tot}$-conserving gate $A(\theta,\phi)$ to generate $|\phi_0^{(r,q)}\rangle$.

In the Heisenberg-chain benchmarks presented here, we use $L_{\rm rand}=N$. Increasing $L_{\rm rand}$ enhances the randomization of the initial state, while for a fixed $L_{\rm rand}$ increasing the number of random samples $R$ can improve the stochastic estimate. Keeping $L_{\rm rand}$ small is advantageous for reducing circuit depth, leading to a trade-off between pseudorandom-state preparation depth and the number of samples. The detailed gate definition, pairing procedure, and the $L_{\rm rand}$ comparison are given in Appendix~\ref{app:circuit}.

\subsection{Finite-shot noise model}
\label{subsec:noise_model}

We introduce
\begin{equation}
\sigma
\equiv
\frac{1}{\sqrt{N_{\mathrm{shot}}}}
\label{eq:sigma}
\end{equation}
as a quantity representing the magnitude of the statistical error arising from a finite number of shots $N_{\mathrm{shot}}$. The value $\sigma=10^{-3}$, primarily used in this study, corresponds to $N_{\mathrm{shot}}=10^{6}$.

In the Hadamard test, the real and imaginary parts of the real-time overlap sequence $g_n$ are obtained as averages over $N_{\mathrm{shot}}$ measurements of $\pm1$ outcomes, with expectation values $\Re g_n$ and $\Im g_n$, respectively. Assuming $N_{\mathrm{shot}}$ is sufficiently large, we approximate this finite-shot statistical error using a Gaussian distribution. Specifically, the real and imaginary parts of the noisy overlap $g_n'$ are modeled as
\begin{align}
\Re g_n'
&\sim
\mathcal{N}\!\left(
\Re g_n,\,
\left(1-(\Re g_n)^2\right)\sigma^2
\right),
\label{eq:noise_real}\\
\Im g_n'
&\sim
\mathcal{N}\!\left(
\Im g_n,\,
\left(1-(\Im g_n)^2\right)\sigma^2
\right),
\label{eq:noise_imag}
\end{align}
where $\mathcal{N}(\mu,v)$ denotes a Gaussian distribution with mean $\mu$ and variance $v$. We do not add noise to the $n=0$ component, as $g_0=1$ is exact. We also impose $g_{-n}'=(g_n')^*$, so that the reconstructed overlap matrix remains Hermitian and Toeplitz.

\begin{figure*}[t]
\centering
\includegraphics[width=\textwidth]{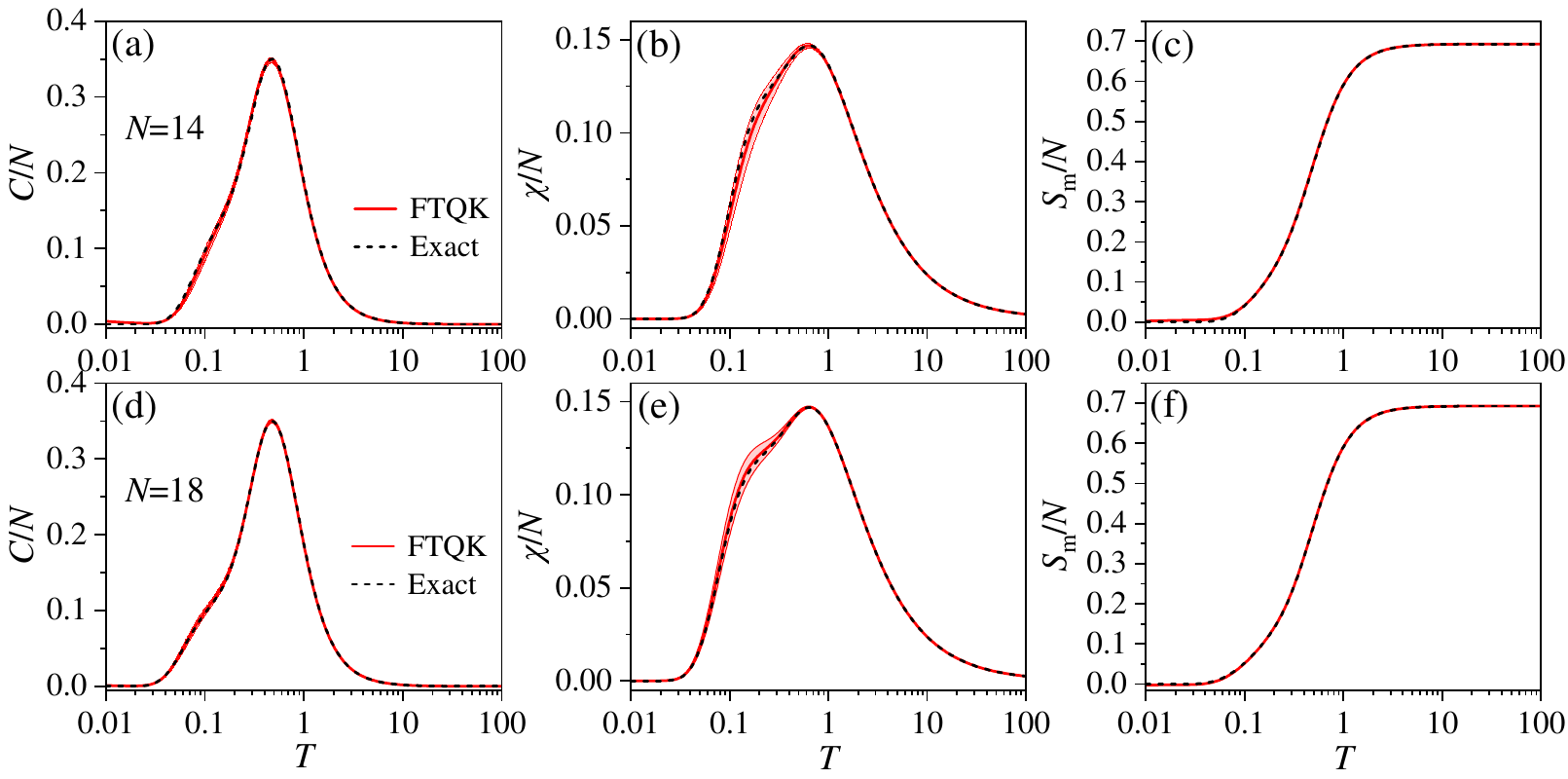}
\caption{
Benchmark results in the absence of noise for the one-dimensional spin-$\frac{1}{2}$ Heisenberg model under PBC. Panels (a)--(c) and (d)--(f) show the specific heat $C/N$, magnetic susceptibility $\chi/N$, and entropy $S_{\mathrm{m}}/N$ for $N=14$ and $N=18$, respectively. The red solid lines show the FTQK results, while the black dashed lines denote the full exact diagonalization results for both system sizes. The shaded regions indicate the standard errors of the FTQK results.
}
\label{F1}
\end{figure*}

\section{Benchmark results}
\label{sec:benchmarks}

\subsection{Noiseless results}
\label{subsec:noiseless_results}

Figure~\ref{F1} shows benchmark results in the absence of noise for the one-dimensional spin-$\frac{1}{2}$ Heisenberg model under PBC. Panels (a)--(c) and (d)--(f) show the specific heat $C/N$, magnetic susceptibility $\chi/N$, and entropy $S_{\mathrm{m}}/N$ for $N=14$ and $N=18$, respectively. For $N=14$, we use $R=100$ and $D=20$, while for $N=18$ we use $R=100$ and $D=40$. Full exact diagonalization is used as the reference for both system sizes.

The thermodynamic curves are assembled sector by sector. For $N=14$, FTQK is applied to the $S^z_{\mathrm{tot}}=0,1,2,3$ sectors and exact diagonalization to the $S^z_{\mathrm{tot}}=4,5,6,7$ sectors. For $N=18$, the same hybrid strategy is used: FTQK is applied to the $S^z_{\mathrm{tot}}=0,1,2,3,4$ sectors and exact diagonalization to the $S^z_{\mathrm{tot}}=5,6,7,8,9$ sectors. Since the low-$|S^z_{\mathrm{tot}}|$ sectors treated by FTQK provide the dominant contributions to the low-temperature thermodynamics, this hybrid comparison directly tests FTQK in the thermodynamically most relevant sectors. 
This sector-resolved treatment also suggests a practical quantum--classical hybrid strategy for future implementations: FTQK can be applied to the low-$\lvert S^z_{\mathrm{tot}}\rvert$ sectors with large Hilbert-space dimensions, while the much smaller high-$\lvert S^z_{\mathrm{tot}}\rvert$ sectors can be treated by exact diagonalization on a classical computer.
For brevity, we refer to the resulting hybrid curves simply as FTQK below.

 As shown in Fig.~\ref{F1}, the FTQK results for both $N=14$ and $N=18$ agree well with those obtained by full exact diagonalization over the plotted temperature range. In particular, because the specific heat is given by the temperature derivative of the internal energy, it is more sensitive to inaccuracies than the internal energy itself. The agreement of the specific heat with the exact result therefore provides a more stringent validation of FTQK than agreement of the internal energy alone. The magnetic susceptibility reproduces its high-temperature behavior, the position and shape of its peak, and its low-temperature decay, while the entropy agrees well with the exact result throughout the plotted temperature range.

\begin{figure}[t]
\centering
\includegraphics[width=0.98\columnwidth]{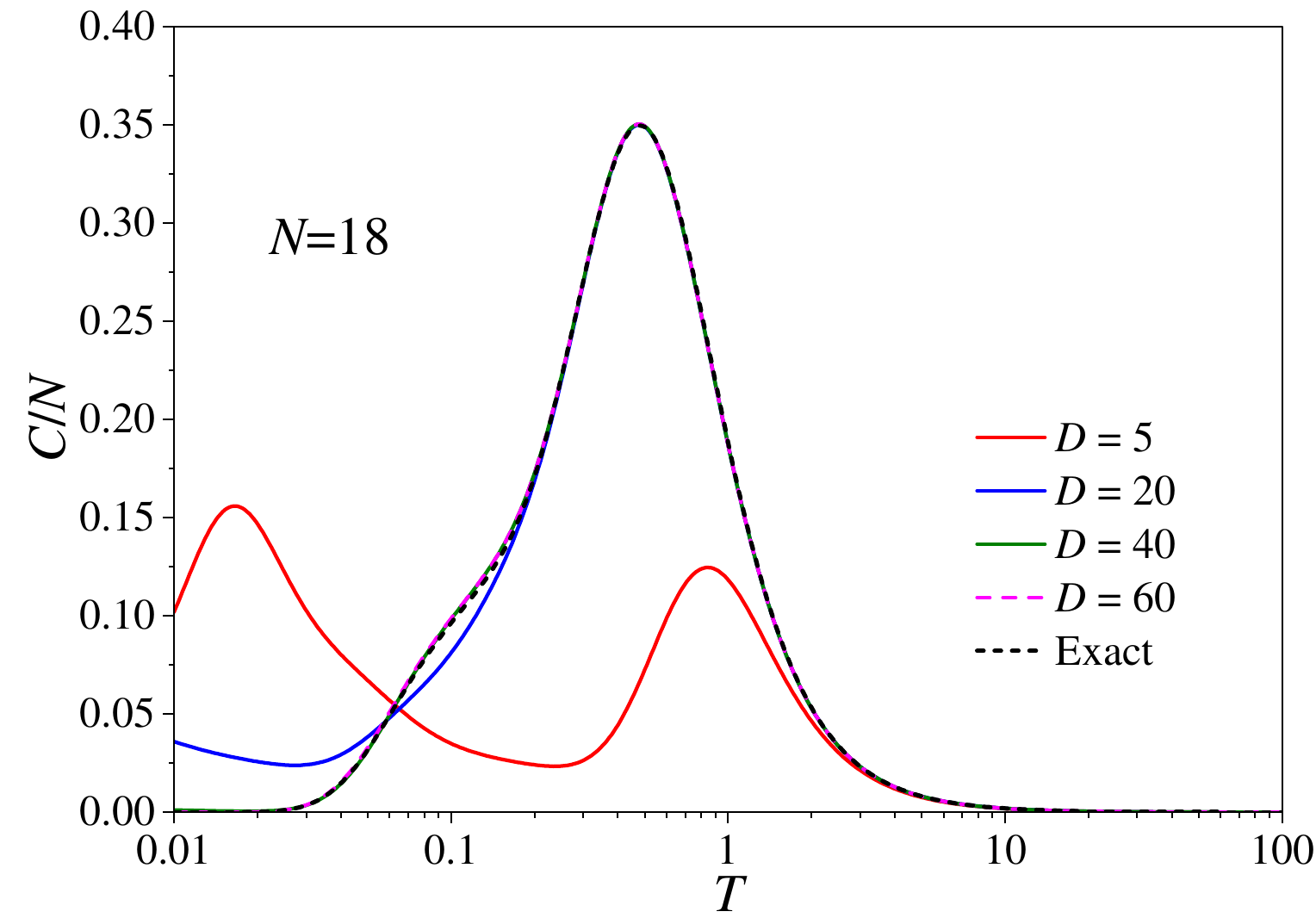}
\caption{Specific heat per site for $N=18$, calculated using $R=100$ random samples with Krylov dimensions $D=5,20,40,$ and $60$. The full exact diagonalization result is shown for comparison. 
The FTQK results systematically converge to the exact result with increasing $D$, and the results for $D=40$ and $60$ are nearly indistinguishable from it.}
\label{fig:D_convergence}
\end{figure}

To examine the convergence with respect to the Krylov dimension $D$, we compare the specific heat for several values of $D$ at $\sigma=0$ in Fig.~\ref{fig:D_convergence}. For small $D$, the FTQK results exhibit substantial deviations from the exact result, particularly at low temperatures. The agreement improves systematically as $D$ increases, and the results for $D=40$ and $60$ are nearly indistinguishable from the exact result over the entire temperature range shown. This demonstrates that highly accurate results can be obtained by choosing a sufficiently large Krylov dimension.

\subsection{Finite-shot-noise results}
\label{subsec:noise_results}

\begin{figure*}[t]
\centering
\includegraphics[width=\textwidth]{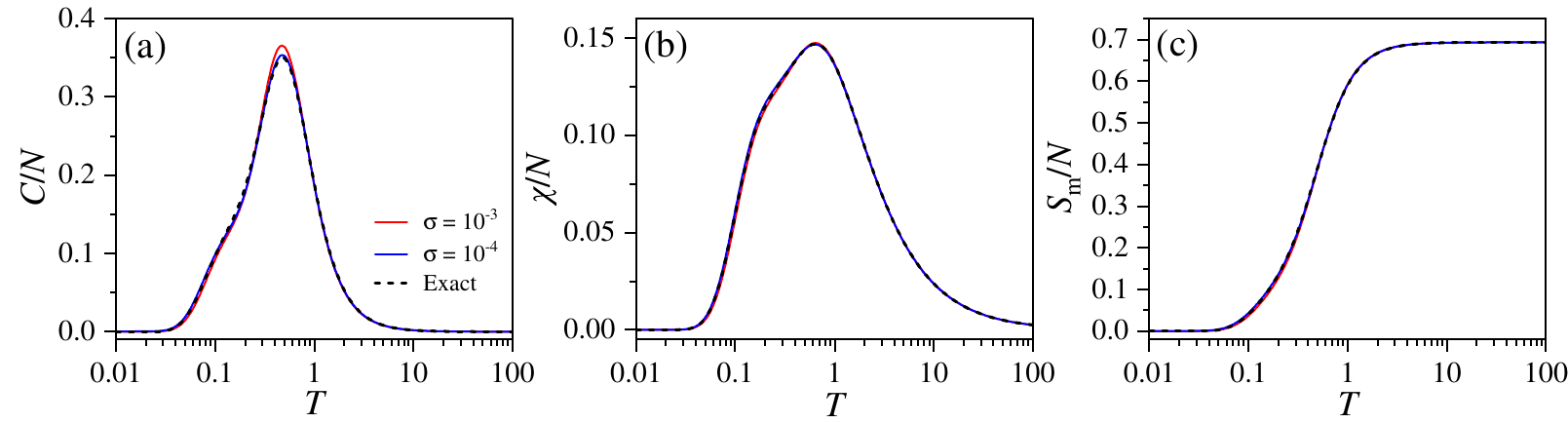}
\caption{Results in the presence of finite-shot noise for the one-dimensional spin-$\frac{1}{2}$ Heisenberg model with $N=14$. Panels (a)--(c) show the specific heat $C/N$, magnetic susceptibility $\chi/N$, and entropy $S_{\mathrm{m}}/N$, respectively. Red and blue lines correspond to $\sigma=10^{-3}$ and $\sigma=10^{-4}$, while the black dashed lines denote the exact reference results. The data are obtained with $R=200$ and $D=50$.}
\label{F2}
\end{figure*}

Next, we examine the behavior of the method under the finite-shot statistical model described in Sec.~\ref{subsec:noise_model}. For the noisy calculations, we regularize the generalized eigenvalue problem through automatic overlap-matrix truncation and additionally stabilize the reconstructed low-energy levels using the trimmed-mean procedure described in Appendix~\ref{app:noise_stability}.

Figure~\ref{F2} shows the results in the presence of finite-shot noise for $N=14$, obtained with $R=200$ and $D=50$. For $\sigma=10^{-4}$, the specific heat, magnetic susceptibility, and entropy are nearly indistinguishable from the reference results on the plotted scale. For $\sigma=10^{-3}$, the main thermodynamic features remain well preserved over a broad temperature range. 
A slight effect of noise is visible in the specific heat around the peak; 
nevertheless, the peak position and the overall shape remain well reproduced in the present data. The magnetic susceptibility and entropy show smaller deviations from the reference results at $\sigma=10^{-3}$.

Appendix~\ref{app:noise_stability} additionally shows that omitting the trimmed-mean stabilization leads to pronounced fluctuations at low temperatures and around the thermodynamic peaks. It also compares the automatic threshold selection with several fixed thresholds, demonstrating the sensitivity of the results to the choice of threshold.

For the larger noise level $\sigma=10^{-2}$, the accuracy deteriorates noticeably, as shown in Appendix~\ref{app:noise_stability}. In practice, compared with the noiseless case, the noisy calculations use larger values of both $R$ and $D$. Stable finite-temperature evaluation under finite-shot noise was achieved here using additional sampling and overlap data together with regularization.

\subsection{Benchmark for the honeycomb Kitaev model}
\label{sec:kitaev_benchmark}

The benchmark calculations presented above focus on the one-dimensional Heisenberg chain, in which the total magnetization $S_{\mathrm{tot}}^z$ is conserved and the thermodynamic quantities can be evaluated sector by sector. To examine the applicability of FTQK beyond simple one-dimensional systems, we next consider the spin-$1/2$ Kitaev model on the honeycomb lattice~\cite{Kitaev2006Honeycomb}. Its two-dimensional geometry and bond-dependent spin interactions provide a qualitatively distinct benchmark for FTQK.
The Hamiltonian is given by
\begin{equation}
H_{\mathrm K}
=
K\sum_{\gamma=x,y,z}
\sum_{\langle i,j\rangle_{\gamma}}
S_i^{\gamma}S_j^{\gamma},
\label{eq:kitaev_hamiltonian}
\end{equation}
where $\langle i,j\rangle_{\gamma}$ denotes a nearest-neighbor bond of type $\gamma$. Throughout this benchmark, we set $K=1$. Since $[H_{\mathrm K},S_{\mathrm{tot}}^z]\neq 0$, the calculation is performed in the full Hilbert space without decomposition into fixed-$S_{\mathrm{tot}}^z$ sectors. Details of the finite cluster, initial-state preparation, and Trotter decomposition are provided in Appendix~\ref{app:additional_model}.

\begin{figure}[t]
\centering
\includegraphics[width=1.0\columnwidth]{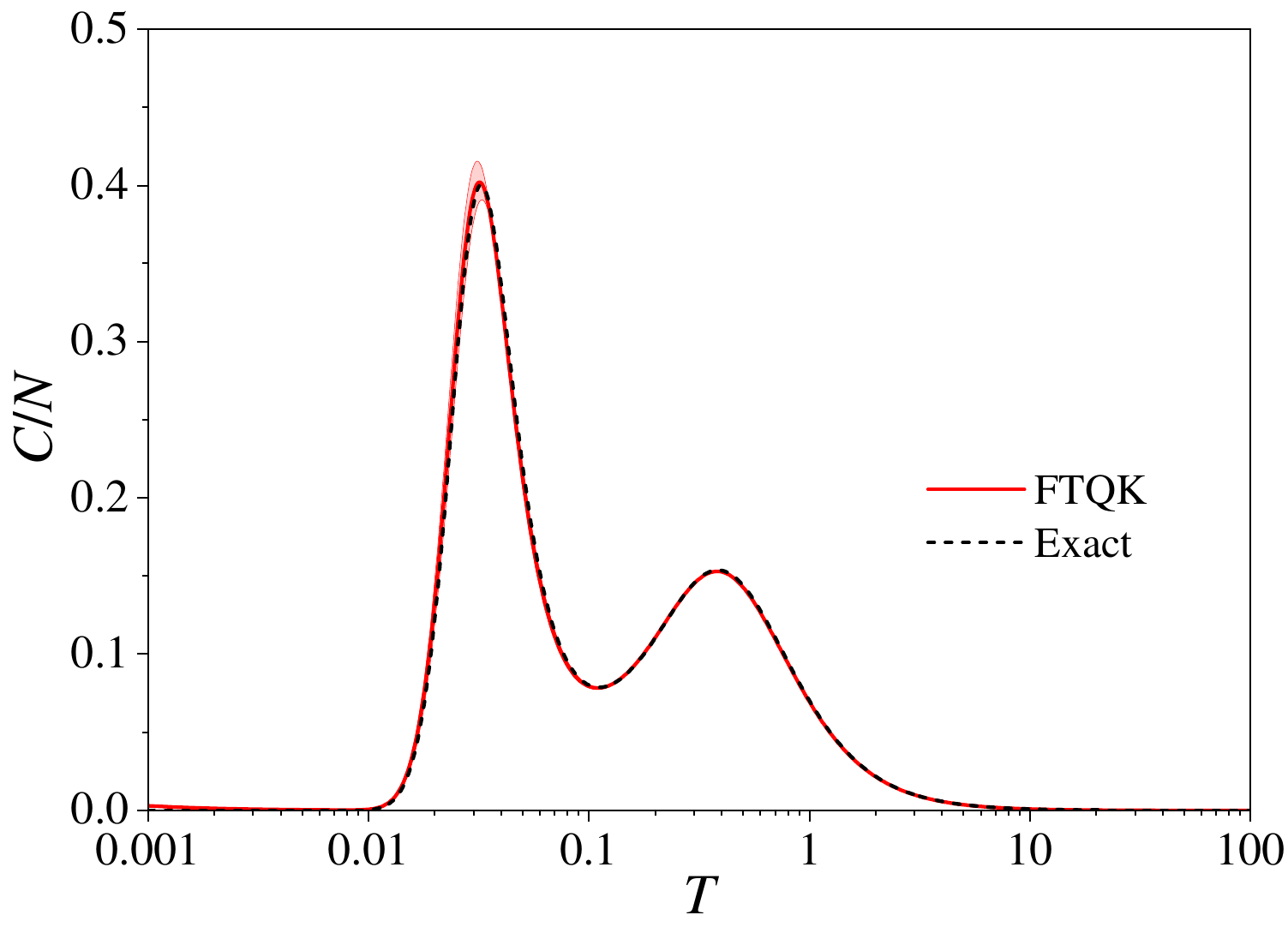}
\caption{Specific heat per site for the $N=14$ honeycomb Kitaev model under PBC. The noiseless FTQK result ($\sigma=0$) is compared with full exact diagonalization. The shaded region indicates the standard error of the FTQK result.
}
\label{fig:kitaev_thermo}
\end{figure}

We perform a noiseless benchmark with $\sigma=0$ for an $N=14$ cluster under PBC, using $R=100$ random samples and a Krylov dimension of $D=50$. Figure~\ref{fig:kitaev_thermo} compares the FTQK result for the specific heat $C/N$ with that obtained by full exact diagonalization. The FTQK result closely agrees with the exact result over the entire temperature range shown. 
This agreement demonstrates that FTQK can accurately reconstruct finite-temperature quantities beyond simple one-dimensional chains, including a two-dimensional model with bond-dependent interactions in which $S_{\mathrm{tot}}^z$ is not conserved.

\section{Discussion}
\label{sec:discussion}

\subsection{Relation to FTLM, QFTLM, and real-time quantum Krylov methods}
\label{subsec:related_methods}

The present construction retains the stochastic trace-estimation logic of FTLM: thermodynamic traces are approximated by averages over random initial states, and spectral information is reconstructed in a low-dimensional Krylov space. In conventional FTLM, the Krylov basis is generated by repeated applications of $H$ to a classical state vector~\cite{Jaklic1994FTLM,Jaklic2000FTLMReview}. In FTQK, the quantum-side information is instead encoded in real-time overlaps generated by unitary evolution. Once these overlaps have been obtained, the generalized eigenvalue problem and the subsequent finite-temperature evaluation are handled entirely by classical post-processing.

The same finite-temperature quantum Krylov method was independently and contemporaneously introduced under the name QFTLM~\cite{Gentinetta2026QFTLM}. Thus, QFTLM and FTQK denote the same method, and the distinction between the two names is purely terminological. Nevertheless, the QFTLM paper and the present study differ in their benchmark models, numerical formulations of the generalized eigenvalue problem, and treatments of finite-shot noise.

First, the study in Ref.~\cite{Gentinetta2026QFTLM} benchmarks the method for the transverse-field Ising model, whereas the present work considers the spin-$1/2$ Heisenberg chain and the honeycomb Kitaev model.
 For the Heisenberg chain, symmetry-preserving pseudorandom states prepared in fixed $S_{\mathrm{tot}}^z$ sectors enable sector-resolved trace estimation and the evaluation of the magnetic susceptibility, in addition to the specific heat and entropy, while the real-time evolution itself is performed in the full Hilbert space. 
The honeycomb Kitaev benchmark in Sec.~\ref{sec:kitaev_benchmark} further extends the validation of FTQK to a two-dimensional model with bond-dependent interactions, for which $S_{\mathrm{tot}}^z$ is not conserved.

Second, the formulation in Ref.~\cite{Gentinetta2026QFTLM} constructs the generalized eigenvalue problem using the projected unitary time-evolution operator $\exp(-i\tau H)$, whereas the present formulation uses the Hermitian operator $\cos(\tilde H)$ after an affine transformation of the Hamiltonian spectrum.
 The reconstructed real eigenvalues are then converted into energies through the inverse-cosine transformation. This distinction becomes particularly important in the presence of finite-shot noise, as discussed below.

Third, the two studies employ different stabilization procedures for finite-shot noise. 
The QFTLM study uses Tikhonov regularization of the Gram matrix, singular-value truncation, and projection onto a unitary matrix~\cite{Gentinetta2026QFTLM}.
In the present work, we examine a finite-shot stabilization procedure tailored to the $\cos(\tilde H)$ formulation. It combines sample-dependent automatic selection of the overlap-matrix threshold $\varepsilon$ with a trimmed-mean treatment of the lowest reconstructed energy levels and their weights. Its ability to mitigate finite-shot errors is demonstrated as a proof of principle for the $N=14$ Heisenberg chain.

In several quantum Krylov approaches, generalized eigenvalue problems based on $H$ or on real-time unitary evolution are considered~\cite{Stair2020MRSQK,Seki2021QuantumPowerMethod,Cortes2022QKSD,Cortes2022FastForwarding,Shen2023RealTimeKrylov}. A direct $H$-based generalized eigenvalue problem requires the Hamiltonian matrix elements $\langle\phi_i|H|\phi_j\rangle$ for all pairs of Krylov basis states. If the time evolution is exact, $U=\exp(-i\tau H)$ commutes with $H$, and these matrix elements depend only on the time difference $j-i$, substantially reducing the number of distinct quantities that must be measured. In a quantum-circuit implementation based on a Trotterized evolution operator $U_{\mathrm T}$, however, $[U_{\mathrm T},H]\neq0$ in general. Consequently, $\langle\phi_i|H|\phi_j\rangle$ depends separately on $i$ and $j$, and up to $D(D+1)/2$ matrix elements in the upper-triangular part must be evaluated individually. The number of Hamiltonian matrix elements to be measured therefore
scales as $O(D^2)$. Moreover, evaluating these matrix elements requires measurements of the terms composing the Hamiltonian, further increasing the measurement overhead.

By contrast, in the $\cos(\tilde H)$-based formulation, both the overlap matrix and the projected cosine matrix can be constructed solely from the ordinary real-time overlap sequence generated by the implemented Trotterized evolution operator. The required number of overlap data therefore remains $O(D)$. This substantial reduction in measurement cost is an important practical advantage of the $\cos(\tilde H)$-based formulation.

The $U$-based formulation extracts energies from the phases of complex eigenvalues. Before unitary projection, as in the example in Appendix~\ref{app:noise_stability}, eigenvalues near the origin can have phases that are highly sensitive to noise. When the phase is close to $\pm\pi$, noise can cause the extracted phase to jump between $-\pi$ and $\pi$. This can make a large negative energy appear as a large positive one, or vice versa, resulting in a substantial error in the reconstructed energy. In Ref.~\cite{Gentinetta2026QFTLM}, unitary projection is used to stabilize the calculation, but this step alone does not guarantee accurate eigenphases.

The present $\cos(\tilde H)$ formulation recovers energies from real eigenvalues through the inverse cosine after an affine spectral mapping. This avoids phase extraction and phase wrapping, and may improve numerical stability under finite-shot noise. However, inverse-cosine recovery is also sensitive to errors near its endpoints.
The general conditioning and sampling-error sensitivity of quantum subspace diagonalization are discussed in Refs.~\cite{Epperly2022TheoryQSD,Kirby2024QKErrors,Lee2024SamplingErrorQKSD,Lee2025ReducingSamplingError}.

Another practical distinction from temperature-dependent state-preparation approaches is that the measured real-time overlaps do not depend on $\beta$. The same quantum data can therefore be reused to evaluate an entire temperature grid in classical post-processing, without preparing a separate thermal state for each target temperature. This reuse is a structural feature of the present method, although it does not imply that the total quantum resource cost is independent of temperature or target accuracy in every implementation.

\subsection{Limitations and outlook}
\label{subsec:limitations}

The choice of the Krylov dimension $D$ involves a trade-off between spectral accuracy and numerical stability. As shown in Fig.~\ref{fig:D_convergence}, a small Krylov dimension produces substantial deviations from the exact specific heat, whereas the noiseless results for $D=40$ and $D=60$ are nearly indistinguishable from the exact result for the $N=18$ benchmark. At the same time, the real-time Krylov basis becomes increasingly linearly dependent as $D$ grows, and small eigenvalues of the overlap matrix amplify numerical and finite-shot errors. The optimal value of $D$ may therefore depend on the system size, model, noise level, and target temperature range. The regularization described in Sec.~\ref{subsec:regularization} is consequently an integral numerical requirement of the generalized eigenvalue problem.

The finite-shot analysis presented here should be regarded as a proof-of-principle study for $N=14$. The calculations at $\sigma=10^{-3}$ use $R=200$ and $D=50$, compared with $R=100$ and $D=20$ in the corresponding noiseless benchmark. In the present implementation, stable results under finite-shot noise were obtained using additional random samples, a larger overlap data set, and regularization. This comparison alone does not disentangle the individual effects of increasing $R$ and $D$, nor does it establish their general resource scaling with system size. In particular, increasing $D$ can improve the spectral reconstruction while simultaneously worsening the conditioning of the overlap matrix. A systematic investigation of the system-size dependence of the automatic-threshold and trimmed-mean stabilization procedures is left for future work. The deterioration observed at $\sigma=10^{-2}$ further limits the conclusions that can be drawn from the present sampling-noise study.

In the symmetry-resolved Heisenberg-chain calculations, the pseudorandom initial states are generated by finite-depth symmetry-preserving circuits rather than by exact Haar sampling. The $L_{\rm rand}$ comparison demonstrates a trade-off between the circuit depth and the number of random samples: increasing the circuit depth improves the mixing of the initial state, while increasing $R$ improves the stochastic trace estimate. However, the present finite-size comparison does not establish the asymptotic typicality properties of the chosen circuit ensemble or the scaling of the required circuit depth and sample number with system size.

Finally, the noise model considered here includes finite-shot statistical fluctuations but not device-specific gate errors, readout errors, decoherence, calibration errors, or hardware-connectivity constraints. The honeycomb Kitaev benchmark in Sec.~\ref{sec:kitaev_benchmark} is also restricted to the noiseless case. 
Broader validation for larger systems and additional interacting models, including finite-shot-noise benchmarks for models without total-magnetization conservation, therefore remains an important direction for future work.
 A hardware implementation will additionally require an assessment of circuit compilation, connectivity, and the accumulation and mitigation of device-level errors.

The present framework is naturally compatible with a future fault-tolerant quantum implementation. In such an implementation, the quantum processor would perform the initial-state preparation, unitary real-time evolution, and overlap measurements, while the generalized eigenvalue problem and thermodynamic reconstruction would be carried out classically. The overlap-based reconstruction could also be combined with classical tensor-network approaches capable of approximating real- or imaginary-time evolution, including density-matrix renormalization group, matrix-product-state, and projected entangled-pair-state methods~\cite{White2004RealTimeDMRG,Stoudenmire2010METTS,Czarnik2012FiniteTPEPS,Lubasch2014FinitePEPS}.

\section{Conclusion}
\label{sec:conclusion}

In this work, we developed FTQK, which reconstructs finite-temperature quantities from temperature-independent real-time overlaps. The method combines stochastic trace estimation with a Hermitian generalized eigenvalue problem based on $\cos(\tilde H)$, while the spectral reconstruction and thermodynamic evaluation are performed entirely by classical post-processing. All benchmark results presented in this work were obtained through classical numerical simulations of the quantum-circuit protocol.

In the noiseless benchmarks, for the one-dimensional spin-$1/2$ Heisenberg model under periodic boundary conditions, the FTQK results for the specific heat, magnetic susceptibility, and entropy agree well with full exact diagonalization for $N=14$ and $N=18$. The Krylov-dimension dependence of the specific heat for $N=18$ shows systematic convergence toward the exact result, with the results for $D=40$ and $D=60$ being nearly indistinguishable from it.
The specific heat obtained for the $N=14$ honeycomb Kitaev model also agrees closely with the exact result, demonstrating that FTQK can be applied beyond simple one-dimensional chains to a two-dimensional model with bond-dependent interactions, even when $S_{\mathrm{tot}}^z$ is not conserved.

For the $N=14$ Heisenberg chain, we further presented a proof-of-principle finite-shot analysis using automatic overlap-matrix truncation and trimmed-mean stabilization. The results remain highly accurate at $\sigma=10^{-4}$, and the main thermodynamic features are preserved at $\sigma=10^{-3}$, whereas the accuracy deteriorates at $\sigma=10^{-2}$. Further improvement of the stabilization procedures and a systematic examination of their system-size dependence are therefore required.

Future work should extend the benchmarks to larger systems and other interacting models. An important next step is to demonstrate the method on quantum hardware for small systems and compare the reconstructed thermodynamic quantities with exact results. In a future implementation on a fault-tolerant quantum computer, the quantum processor would perform the initial-state preparation, unitary real-time evolution, and overlap measurements, while the generalized eigenvalue problem and thermodynamic reconstruction would be handled classically. Because the real-time overlaps are independent of temperature, the same quantum data can be reused across the entire temperature grid.

\acknowledgments{The authors thank H. Ueda and N. Shibata for valuable discussions.
This work was supported by JSPS KAKENHI Grant No. JP25K07230.}

\section*{Author contributions}
\label{sec:author_contributions}
H.Y. performed the numerical calculations, refined the method, and compiled the numerical data. 
K.M. conceived the method, supervised the project, and prepared the figures. Both authors contributed to drafting and revising the manuscript.
ChatGPT (OpenAI) was used to assist with English-language editing, literature searches, and checks of mathematical and textual consistency. The authors reviewed and revised the AI-assisted material and take full responsibility for the content of the manuscript.

\bibliographystyle{quantum}
\bibliography{ref_Quantum}

\newpage

\appendix

\section{Heisenberg-chain pseudorandom-state preparation}
\label{app:circuit}

\subsection{Symmetry-preserving \texorpdfstring{$A$}{A} gate}
\label{appsub:A_gate}

\begin{figure}[H]
\centering
\includegraphics[width=0.95\columnwidth]{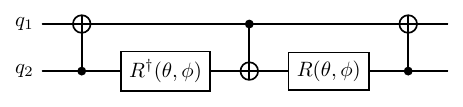}
\caption{Quantum circuit decomposition of the $A(\theta,\phi)$ gate up to a global phase. The $A(\theta,\phi)$ gate is a two-qubit gate that conserves the total magnetization $S^z_{\rm tot}$ and can be decomposed into three CNOT gates and two single-qubit rotations $R(\theta,\phi)$.}
\label{fig:Agate}
\end{figure}

For the Heisenberg-chain benchmark, the $A$ gate is employed as a two-qubit gate that conserves the total magnetization $S^z_{\mathrm{tot}}$. Following the symmetry-preserving state-preparation circuit design of Gard \textit{et al.}~\cite{Agate}, we define $A(\theta,\phi)$ in the computational basis
$\{|00\rangle,|01\rangle,|10\rangle,|11\rangle\}$ as
\begin{equation}
A(\theta,\phi)
=
\begin{pmatrix}
1 & 0 & 0 & 0 \\
0 & \cos\theta & e^{i\phi}\sin\theta & 0 \\
0 & e^{-i\phi}\sin\theta & -\cos\theta & 0 \\
0 & 0 & 0 & 1
\end{pmatrix}.
\label{eq:A_gate}
\end{equation}
The angles are independently sampled from the uniform distributions
\begin{equation}
\theta \sim \mathrm{Uniform}[0,\pi],
\qquad
\phi \sim \mathrm{Uniform}[0,2\pi).
\label{eq:A_gate_random_angles}
\end{equation}
Because this gate mixes states only within the subspace spanned by $|01\rangle$ and $|10\rangle$, it conserves the total magnetization of the two target qubits. Furthermore, the $A$ gate can be expressed in terms of Pauli operators as
\begin{equation}
\begin{aligned}
A(\theta,\phi)
={}&e^{i\pi/4}
e^{+\frac{i}{2}\left(\frac{\pi}{2}-\phi\right)Z_2}
e^{-\frac{i}{2}\left(\theta(X_1X_2+Y_1Y_2)+\frac{\pi}{2}Z_1Z_2\right)}\\
&\times e^{-\frac{i}{2}\left(\frac{\pi}{2}Z_1-\phi Z_2\right)}.
\end{aligned}
\label{eq:A_gate_pauli}
\end{equation}
For quantum-device implementation, this $A$ gate can be decomposed into three CNOT gates and two single-qubit rotations $R(\theta,\phi)$, as shown in Fig.~\ref{fig:Agate}. Here,
$R(\theta,\phi)=R_z(\phi+\pi)R_y(\theta+\pi/2)$,
where $R_z(\alpha)=\exp(-i\alpha Z/2)$ and
$R_y(\alpha)=\exp(-i\alpha Y/2)$.

\subsection{Pseudorandom-state preparation for the Heisenberg chain}
\label{appsub:pseudorandom_circuit}

The initial state $|\phi_0^{(r,q)}\rangle$ used in the Heisenberg-chain benchmark is not a Haar-random state in the $S^z_{\mathrm{tot}}=q$ sector, but rather a pseudorandom state generated by a finite-depth random circuit that conserves the total magnetization. In the following, we assume $q\ge0$. Setting
$n_\downarrow=N/2-q$ and $n_\uparrow=N/2+q$, we first randomly select a computational basis state $|c^{(r,q)}\rangle$ containing $n_\downarrow$ spins in the $\ket{1}$ state and $n_\uparrow$ spins in the $\ket{0}$ state. We then apply $L_{\rm rand}$ layers of the two-qubit gate $A(\theta,\phi)$, with independently sampled random angles, to generate $|\phi_0^{(r,q)}\rangle$. The specific procedure is as follows:
\begin{itemize}
\item $L_{\rm rand}=0$: We prepare $|c^{(r,q)}\rangle$ as the initial state.
\item $L_{\rm rand}=1$: The first layer randomly forms $n_\downarrow$ non-overlapping pairs between qubits in the $\ket{1}$ state and qubits in the $\ket{0}$ state, and applies $A(\theta,\phi)$ to each pair.
\item $L_{\rm rand}=2$: The second layer randomly pairs the $\ket{0}$-state qubits that were not acted upon in the first layer with qubits already acted upon, and applies $A(\theta,\phi)$ to each pair. For $q=0$, no additional operations occur in this layer; nevertheless, it is counted as one layer for consistency with the other magnetization sectors.
\item $L_{\rm rand}\ge3$: Subsequent layers randomly form non-overlapping pairs from all qubits and apply $A(\theta,\phi)$ to each pair.
\end{itemize}

\begin{figure*}[t]
\centering
\includegraphics[width=0.92\textwidth]{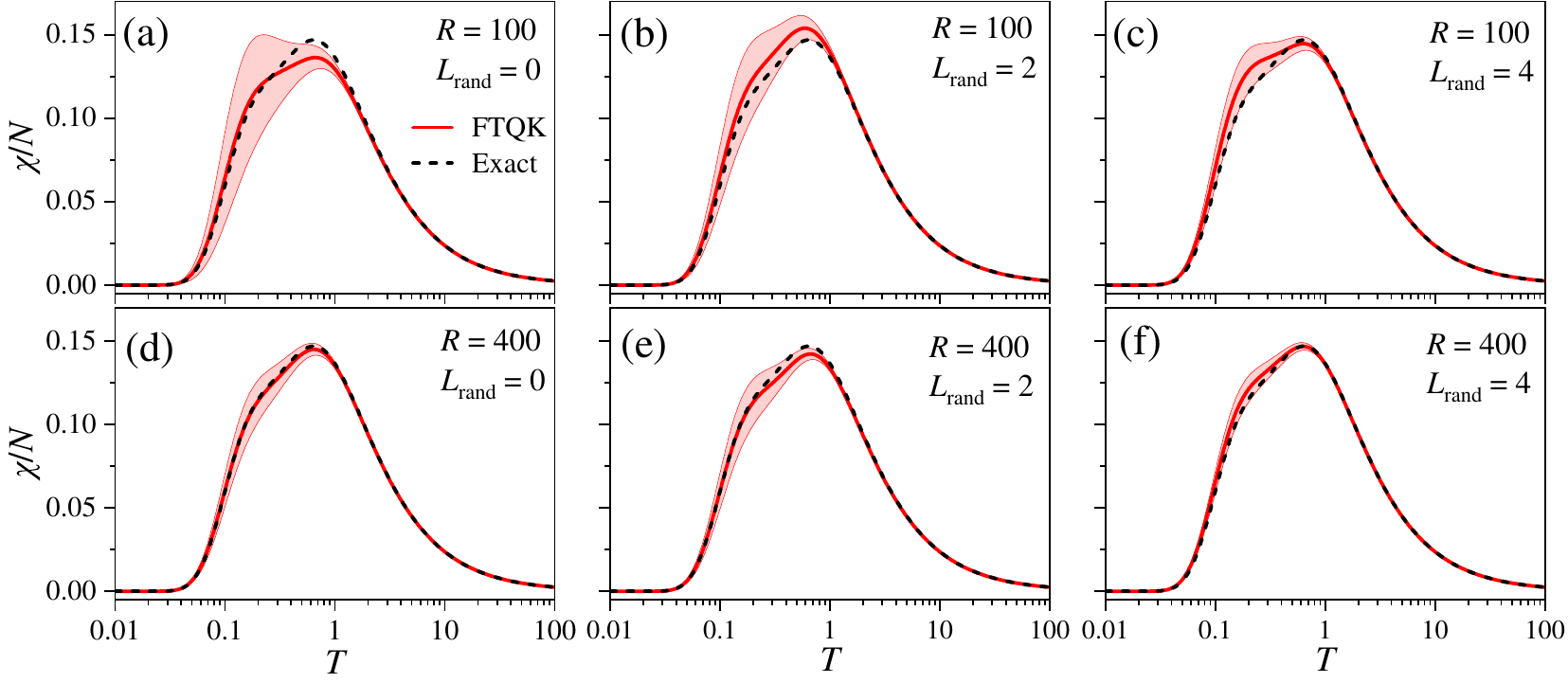}
\caption{Dependence of the magnetic susceptibility on the number of layers $L_{\rm rand}$ in the pseudorandom-state preparation circuit for $N=14$, $D=20$, and $\sigma=0$. The shaded regions indicate the standard errors over the random samples. Increasing $L_{\rm rand}$ enhances the mixing of the initial states and tends to improve the agreement with the exact result. For the displayed values of $L_{\rm rand}$, increasing the number of random samples from $R=100$ to $R=400$ also improves the agreement.}
\label{fig:Lrand}
\end{figure*}

Through this procedure, pseudorandom states are generated while remaining within the fixed $S^z_{\mathrm{tot}}=q$ sector. In the Heisenberg-chain benchmarks presented in the main text, we use $L_{\rm rand}=N$. As shown in Fig.~\ref{fig:Lrand}, increasing $L_{\rm rand}$ enhances the mixing of the initial states and tends to improve the agreement with the exact result. For fixed $L_{\rm rand}$, increasing the number of random samples $R$ also improves the stochastic estimate. Thus, reducing the state-preparation depth by using a smaller $L_{\rm rand}$ leads to a trade-off with the number of random samples required.

\FloatBarrier

\section{Stability analysis under finite-shot noise}
\label{app:noise_stability}

\subsection{Automatic threshold selection}
\label{appsub:auto_threshold}

The regularization itself is defined in Sec.~\ref{subsec:regularization}. To eliminate nonphysical eigenvalues caused by noise and numerical errors in the present implementation, we determine the threshold $\varepsilon$ using the reference value for the spectral lower bound $E_{\min,\mathrm{ref}}^{(q)}$ introduced in Eq.~\eqref{eq:Eref}.

Specifically, based on the eigenvalues of the overlap matrix $S$, we decrement the effective subspace dimension $D_{\mathrm{eff}}$ one by one from $D$ and solve the standard eigenvalue problem in the reduced subspace at each step. We recover the energies $E_j$ from the obtained eigenvalues using Eq.~\eqref{eq:E_recover_final}, and adopt the threshold at the point where the minimum recovered energy becomes larger than the reference value $E_{\min,\mathrm{ref}}^{(q)}$.

Because an excessively small $\varepsilon$ amplifies noise and rounding errors, we introduce an absolute lower bound $\varepsilon_{\mathrm{floor}}$. In the presence of noise, thresholds comparable to the sampling-error scale are commonly used in quantum subspace regularization~\cite{Epperly2022TheoryQSD,Kirby2024QKErrors}. For the noisy calculations, we set
$\varepsilon_{\mathrm{floor}}=0.1\,\sigma$
for $\sigma=10^{-2},10^{-3}$, and $10^{-4}$.
For the noiseless calculations $(\sigma=0)$, the automatic
threshold-selection procedure is not applied; instead, we use a fixed
numerical cutoff $\varepsilon=10^{-14}$ to remove numerically negligible
eigenvalues of the overlap matrix.
The automatic threshold-selection procedure used for the noisy calculations
is summarized in Algorithm~\ref{alg:auto_threshold_dim}.

\begin{algorithm}[H]
\caption{Automatic threshold selection}
\label{alg:auto_threshold_dim}
\begin{algorithmic}[1]
\State input $(s_i)_{i=1}^{D}$, $s_i\ge s_{i+1}$, $E_{\mathrm{ref,min}}$, $\sigma$
\For{$D_{\mathrm{eff}}=D,D-1,\dots,1$}
    \If{$s_{D_{\mathrm{eff}}}<\varepsilon_{\mathrm{floor}}$}
        \State \textbf{continue}
    \EndIf
    \State $\varepsilon\gets s_{D_{\mathrm{eff}}}$
    \State $F_\varepsilon\gets X_\varepsilon^\dagger F X_\varepsilon$
    \State $(\lambda_j)_{j=1}^{D_{\mathrm{eff}}}=\mathrm{eigvals}(F_\varepsilon)$
    \State recover $(E_j)_{j=1}^{D_{\mathrm{eff}}}$ from Eq.~\eqref{eq:E_recover_final}
    \State $E_{\min}\gets\min_j E_j$
    \If{$E_{\min}>E_{\mathrm{ref,min}}$}
        \State return $\varepsilon_{\mathrm{auto}}\gets\varepsilon$
    \EndIf
\EndFor
\State return $\varepsilon_{\mathrm{auto}}\gets\varepsilon_{\mathrm{floor}}$
\end{algorithmic}
\end{algorithm}

\begin{figure*}[t]
\centering
\includegraphics[width=\textwidth]{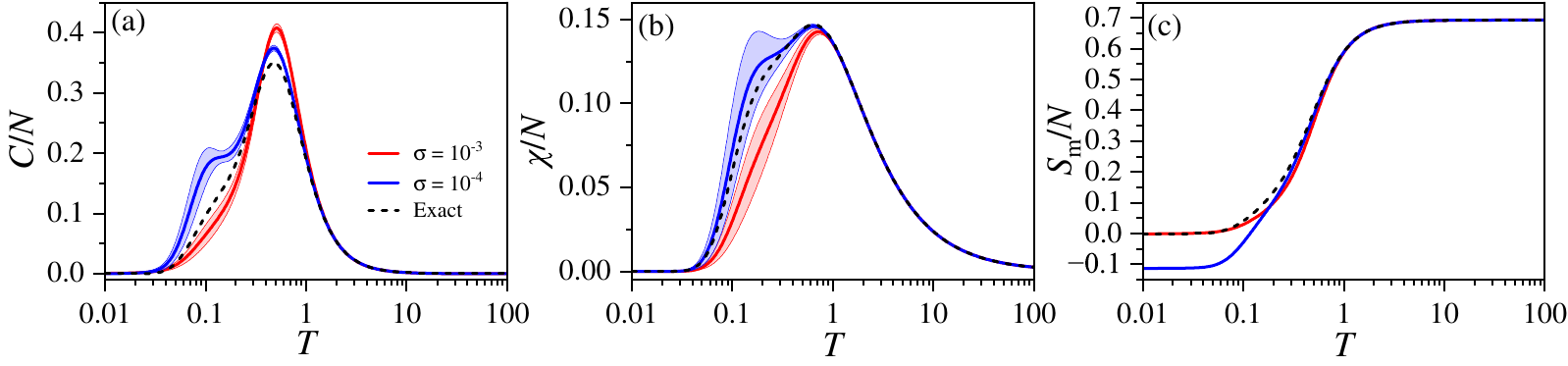}
\caption{Temperature dependence of thermodynamic quantities when automatic threshold selection is applied without stabilization of the low-energy levels via the trimmed mean. Simulations were performed for $N=14$, $R=200$, and $D=50$. Without the trimmed-mean correction, deviations from the exact solutions due to finite-shot noise are visible, especially from the low-temperature region up to the peak positions of the specific heat and magnetic susceptibility.}
\label{fig:no_trimmed}
\end{figure*}

\subsection{Low-energy stabilization via trimmed mean}
\label{appsub:trimmed_mean}

Due to noise, the approximate eigenvalues and weights can fluctuate significantly across samples. These fluctuations severely destabilize the evaluation of thermodynamic quantities in the low-temperature regime. To mitigate this, the present implementation introduces a stabilization procedure that assigns common representative values to the ground and low-lying excited states across all samples.

Specifically, for the two low-energy target levels, namely the ground state ($k=0$) and the first-excited state ($k=1$), let $E_k^{(r)}$ denote the approximate energy obtained from the $r$th sample. To reduce the influence of outliers, the representative energy is defined using the central $50\%$ trimmed mean,
\begin{equation}
\widetilde E_k
=
\operatorname{TrimMean}_{50\%}
\left(
E_k^{(1)},E_k^{(2)},\ldots,E_k^{(R)}
\right).
\label{eq:trimmed_energy}
\end{equation}
The corresponding energies in all samples are then replaced by $\widetilde E_k$. Similarly, representative values for the weights of these two low-energy levels are determined using the central $50\%$ trimmed mean.

Letting $\bar w_k$ denote the trimmed mean of the weight for the $k$th target level, we use the relation
$\bar w_k\simeq\nu_k/N_{\mathrm{st}}^{(q)}$,
where $\nu_k$ is the degeneracy of the $k$th level. An estimated effective degeneracy $\nu_k^{\mathrm{est}}$ is obtained by rounding
$\bar w_kN_{\mathrm{st}}^{(q)}$
to the nearest integer, and the corresponding weight is then fixed to
\begin{equation}
\widetilde w_k
=
\frac{\nu_k^{\mathrm{est}}}{N_{\mathrm{st}}^{(q)}},
\qquad k=0,1
\label{eq:trimmed_weight}
\end{equation}

After fixing the approximate energies and weights of the low-energy levels, the weights of the remaining higher-energy states are uniformly rescaled so that the total weight is unity in each sample. Specifically, for $j\ge 2$,
\begin{equation}
\widetilde w_j^{(r)}
=
\alpha^{(r)} w_j^{(r)},
\qquad
\alpha^{(r)}
=
\frac{
1-\widetilde w_0-\widetilde w_1
}{
\sum_{j=2}^{D_{\mathrm{eff}}-1} w_j^{(r)}
}.
\label{eq:weight_rescaling}
\end{equation}
This normalization ensures the correct infinite-temperature normalization of the spectral weights, and consequently of the partition function and entropy. This procedure is applied to the noisy data only.

\subsection{Additional finite-shot-noise results}
\label{appsub:noise_ablation}
Figure~\ref{fig:no_trimmed} shows the results when automatic threshold selection is applied without stabilizing the low-energy levels via the trimmed mean. Under this condition, large fluctuations caused by finite sampling noise are observed, particularly from the low-temperature regime to around the peaks of the specific heat and magnetic susceptibility.

\begin{figure*}[t]
\centering
\includegraphics[width=\textwidth]{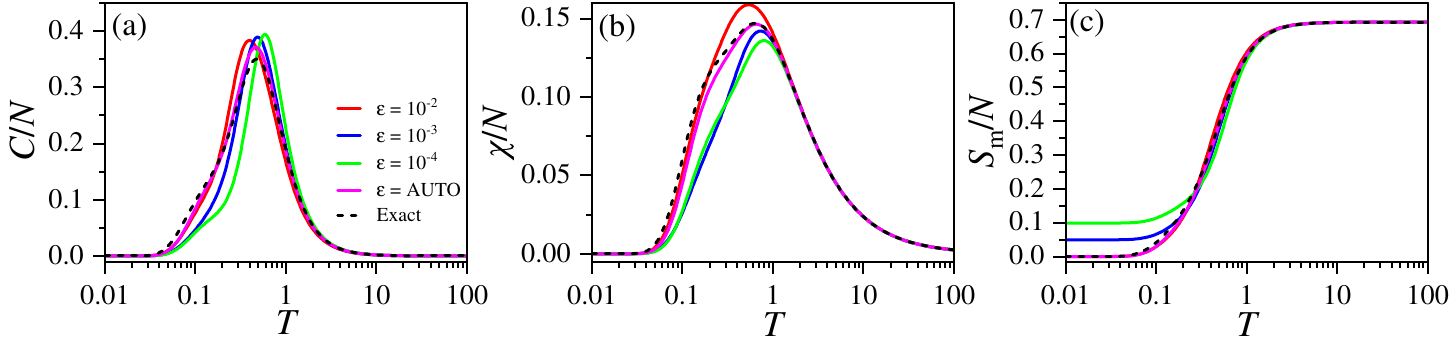}
\caption{Comparison of automatic threshold selection with fixed thresholds for $N=14$, $R=200$, $D=50$, and $\sigma=10^{-3}$. The trimmed mean is applied in all calculations. The same Gaussian noise sequence is used for the fixed thresholds $\varepsilon=10^{-2},10^{-3},10^{-4}$ and the automatic threshold (AUTO).}
\label{fig:no_auto_eps}
\end{figure*}

Figure~\ref{fig:no_auto_eps} compares automatic threshold selection with fixed thresholds $\varepsilon=10^{-2},10^{-3}$, and $10^{-4}$. The trimmed mean is applied in all calculations, and the identical Gaussian noise sequence is used across all conditions in the comparison.

\begin{figure*}[t]
\centering
\includegraphics[width=\textwidth]{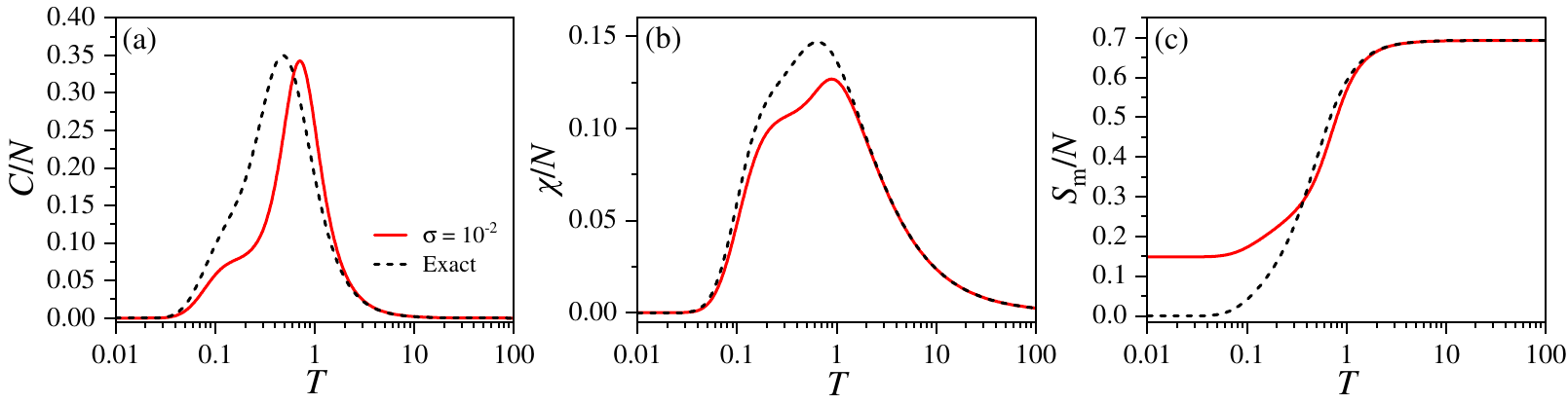}
\caption{Calculation results at $\sigma=10^{-2}$ for the one-dimensional spin-$\frac{1}{2}$ Heisenberg model with $N=14$, $R=200$, and $D=50$. Both the trimmed mean and automatic threshold selection are applied. Deviations from the exact solution are visible near the peaks of the specific heat and magnetic susceptibility and in the low-temperature entropy.}
\label{fig:sigma10m2}
\end{figure*}

With the larger finite-shot noise $\sigma=10^{-2}$, deviations from the exact solutions become more apparent. Figure~\ref{fig:sigma10m2} shows the result with both the trimmed mean and automatic threshold selection. Discrepancies appear in the peak positions and shapes of the specific heat and magnetic susceptibility, and errors are also visible in the low-temperature entropy.

\FloatBarrier

\subsection{Generalized eigenvalue problem based on $U(\tau)$}
\label{appsub:U_based}
\begin{figure}[H]
\centering
\includegraphics[width=0.75\columnwidth]{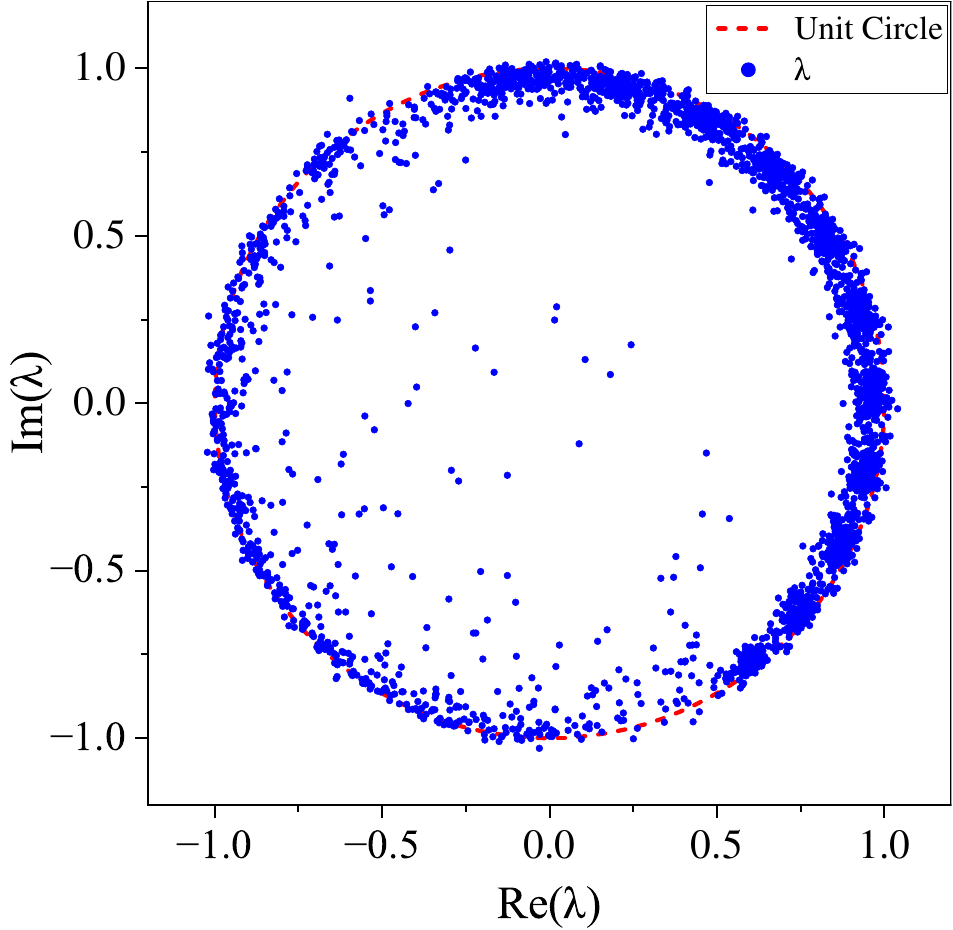}
\caption{Example of complex Ritz values obtained from the generalized eigenvalue problem directly using $U(\tau)=e^{-i\tau H}$ without applying a projection onto a unitary matrix. The parameters are $N=14$, $D=20$, $\varepsilon=10^{-3}$, $\sigma=10^{-3}$, $R=200$, $S^z_{\mathrm{tot}}=0$, and $\tau=dt=0.3$.}
\label{fig:unit_circle}
\end{figure}

In the main text, we use the generalized eigenvalue problem based on the Hermitian operator $\cos(\tilde H)$. Here, we consider the generalized eigenvalue problem based directly on the time-evolution operator $U(\tau)=e^{-i\tau H}$.
Although $U(\tau)$ itself is unitary, its projection onto a finite Krylov subspace is not generally unitary. Therefore, the Ritz values obtained from the generalized eigenvalue problem do not necessarily lie on the unit circle even in the absence of sampling noise. In the $U(\tau)$-based formulation, the approximate energies are obtained from the phases of the complex Ritz values.

Figure~\ref{fig:unit_circle} shows an example of the complex Ritz values under finite-shot noise without applying a projection onto a unitary matrix. Ritz values near the origin can have phases that are highly sensitive to noise, and phase wrapping near $\pm\pi$ can also lead to large errors in the reconstructed energies. This observation motivates the Hermitian formulation based on $\cos(\tilde H)$ adopted here.

\clearpage

\section{Honeycomb Kitaev-model setup}
\label{app:additional_model}

\subsection{Finite cluster and Trotter decomposition}
\label{appsub:kitaev_cluster}

\begin{figure}[H]
\centering
\includegraphics[width=0.98\columnwidth]{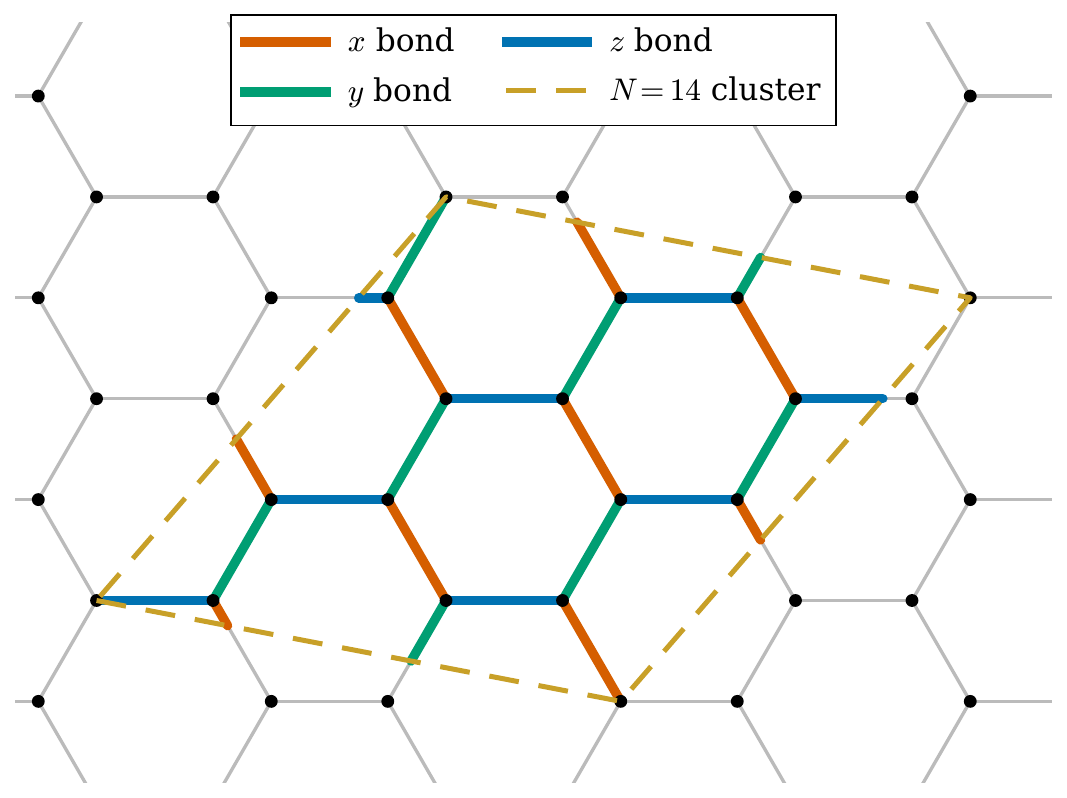}
\caption{Finite $N=14$ honeycomb cluster with PBC used for the Kitaev-model benchmark. The cluster preserves $C_6$ rotational symmetry. The colored bonds indicate the $x$-, $y$-, and $z$-bond groups used in the Trotter decomposition in Eq.~\eqref{eq:kitaev_trotter}.}
\label{fig:kitaev_cluster}
\end{figure}

For the honeycomb Kitaev benchmark introduced in Sec.~\ref{sec:kitaev_benchmark}, we use an $N=14$ finite cluster under PBC chosen to preserve the $C_6$ rotational symmetry, as shown in Fig.~\ref{fig:kitaev_cluster}. For the Trotter decomposition, the Hamiltonian in Eq.~\eqref{eq:kitaev_hamiltonian} is divided into the three bond groups as
\begin{equation}
H_{\mathrm K}=H_x+H_y+H_z.
\end{equation}
A second-order Trotter step is then given by
\begin{equation}
e^{-i\,dt H_{\mathrm K}}
\simeq
e^{-i\frac{dt}{2}H_x}
e^{-i\frac{dt}{2}H_y}
e^{-i\,dt H_z}
e^{-i\frac{dt}{2}H_y}
e^{-i\frac{dt}{2}H_x}.
\label{eq:kitaev_trotter}
\end{equation}
Within each bond group, the two-site terms act on non-overlapping bonds and can therefore be applied in parallel. We use \(N_{\mathrm{step}}=1\) in the present benchmark, where \(N_{\mathrm{step}}\) denotes the number of Trotter steps used to approximate each time-evolution operator. For the affine spectral mapping, the exact spectral edges of the $N=14$ cluster are extended by a $5\%$ margin on both sides:
\begin{equation*}
\begin{aligned}
E_{\min,\mathrm{ref}}
&=
E_{\min,\mathrm{exact}}
-
0.05|E_{\min,\mathrm{exact}}|,\\
E_{\max,\mathrm{ref}}
&=
E_{\max,\mathrm{exact}}
+
0.05|E_{\max,\mathrm{exact}}|.
\end{aligned}
\end{equation*}

\subsection{Pseudorandom-state preparation}
\label{appsub:kitaev_random_state}

Since the Kitaev Hamiltonian does not conserve $S^z_{\mathrm{tot}}$, the initial states are not restricted to fixed-$S^z_{\mathrm{tot}}$ sectors but are generated in the full $2^N$-dimensional Hilbert space. For each random sample $r$, we construct a pseudorandom state by applying $L$ layers of a parameterized random circuit to the product state $|0\rangle^{\otimes N}$~\cite{Sim2019Expressibility}. We denote the unitary operator corresponding to the $\ell$th layer by $U_{\mathrm{rand}}^{(r,\ell)}$.

We label the qubits by $i=0,\ldots,N-1$. Each random-circuit layer consists of two successive stages. In the first stage, independent single-qubit rotations $R_y(\theta)$ and $R_z(\phi)$ are applied to all qubits. These rotations are followed by CNOT gates acting on the non-overlapping nearest-neighbor pairs
\begin{equation}
(0,1),\ (2,3),\ (4,5),\ \ldots .
\end{equation}
In the second stage, independent $R_y(\theta)$ and $R_z(\phi)$ rotations are applied again, now to the internal qubits $i=1,\ldots,N-2$, followed by CNOT gates acting on the shifted non-overlapping pairs
\begin{equation}
(1,2),\ (3,4),\ (5,6),\ \ldots .
\end{equation}
Thus, neighboring qubits are coupled in an alternating pattern within each layer, allowing the random circuit to spread correlations throughout the system as the number of layers increases.

All rotation angles appearing at different qubits, stages, layers, and random samples are sampled independently from
\begin{equation}
\theta,\phi\sim\mathrm{Uniform}[0,2\pi).
\label{eq:kitaev_random_angles}
\end{equation}
The resulting pseudorandom initial state is
\begin{equation}
|\phi_0^{(r)}\rangle
=
U_{\mathrm{rand}}^{(r,L)}
\cdots
U_{\mathrm{rand}}^{(r,2)}
U_{\mathrm{rand}}^{(r,1)}
|0\rangle^{\otimes N}.
\label{eq:kitaev_random_angles}
\end{equation}
For the $N=14$ honeycomb Kitaev benchmark, we use $L=N=14$ layers and $R=100$ independently generated pseudorandom states.\end{document}